%% file: 0_main.tex
\documentclass[sigconf]{acmart}

\input{setting_packages}

\input{setting_variable}

\input{setting_fixed}

\AtBeginDocument{%
  \providecommand\BibTeX{{%
    \normalfont B\kern-0.5em{\scshape i\kern-0.25em b}\kern-0.8em\TeX}}}

\copyrightyear{2026}
\acmYear{2026}
\setcopyright{cc}
\setcctype{by}
\acmConference[MSR ’26]{23rd International Conference on Mining Software Repositories}{April 13--14, 2026}{Rio de Janeiro, Brazil}
\acmBooktitle{23rd International Conference on Mining Software Repositories (MSR ’26), April 13--14, 2026, Rio de Janeiro, Brazil}
\acmPrice{}
\acmDOI{10.1145/3793302.3793591}
\acmISBN{979-8-4007-2474-9/2026/04}

\acmConference[MSR 2026]{MSR '26: Proceedings of the 23rd International Conference on Mining Software Repositories}{April 2026}{Rio de Janeiro, Brazil}
\acmISBN{978-1-4503-XXXX-X/18/06}

\begin{document}

\title{What to Cut? Predicting Unnecessary Methods\\ in Agentic Code Generation}
\renewcommand{\shorttitle}{What to Cut? Predicting Unnecessary Methods in Agentic Code Generation}

\author{
  Kan Watanabe\textsuperscript{$\dagger$},
  Tatsuya Shirai\textsuperscript{$\dagger$},
  Yutaro Kashiwa\textsuperscript{$\dagger$},
  Hajimu Iida\textsuperscript{$\dagger$}
}

\affiliation{%
  \institution{\textsuperscript{$\dagger$}Nara Institute of Science and Technology, Japan}
  \country{}
}

\email{{watanabe.kan.wi6, shirai.tatsuya.sp1, yutaro.kashiwa, iida}@naist.ac.jp}





\renewcommand{\shortauthors}{Watanabe, et al.}

\begin{abstract}
Agentic Coding, powered by autonomous agents such as GitHub Copilot and Cursor, enables developers to generate code, tests, and pull requests from natural language instructions alone. While this accelerates implementation, it produces larger volumes of code per pull request, shifting the burden from implementers to reviewers. In practice, a notable portion of AI-generated code is eventually deleted during review, yet reviewers must still examine such code before deciding to remove it. No prior work has explored methods to help reviewers efficiently identify code that will be removed.

In this paper, we propose a prediction model that identifies functions likely to be deleted during PR review. Our results show that functions deleted for different reasons exhibit distinct characteristics, and our model achieves an AUC of 87.1\%. These findings suggest that predictive approaches can help reviewers prioritize their efforts on essential code.
\end{abstract}



\maketitle

\input{1_Introduction}
\input{2_Background}
\input{3_StudyDesign}

\input{4_RQ1}
\input{4_RQ3}

\input{6_RelatedWork}

\input{7_Threat2Validity}

\section{Conclusion and Future work}\label{sec:conclusion}
To address the decline in code maintainability and the increased review load associated with the proliferation of Agentic Coding tools such as GitHub Copilot and Cursor, this study empirically investigated the characteristics of AI-generated methods deleted during code review and built a prediction model to identify methods likely to be deleted at the time of pull request creation.

Our results reveal that 9.9\% of methods created by Agentic Coding are deleted in environments where method deletion occurs.
Furthermore, the deleted methods tended to exhibit redundancy across their function names, code content, and documentation. 
Finally, our prediction model demonstrated high classification performance, achieving an average AUC of 87.1\%.

In future work, we plan to extend our approach by incorporating additional metrics that detect redundant methods within the existing code base, such as semantic similarity with other methods and duplicate functionality detection, to further improve prediction accuracy and support reviewers in identifying unnecessary code.


\begin{acks}
This work was supported by JSPS KAKENHI (JP24K02921, JP25K21359) and JST PRESTO (JPMJPR22P3), ASPIRE (JPMJAP2415), and AIP Accelerated Program (JPMJCR25U7).

\end{acks}

\balance
\bibliographystyle{ACM-Reference-Format}
\bibliography{references}


\end{document}

%% file: setting_packages.tex
\usepackage{fancyvrb}
\usepackage{xurl}
\usepackage{booktabs}
\usepackage{url}
\usepackage{amsmath,amsfonts}
\usepackage{algorithmic}
\usepackage{graphicx}
\usepackage{textcomp}
\usepackage{xspace}
\usepackage{subcaption}
\usepackage{tabularx}
\usepackage{framed}
\usepackage{chngpage}
\usepackage[many]{tcolorbox}
\usepackage{pgfplots}
\pgfplotsset{compat=newest}
\usepackage{siunitx}
\usepackage{balance}
\usepackage{enumitem}
\usepackage{listings} 
\usepackage{multirow} 
\usepackage{xfp}      

\usepackage{hyperref}

\usepackage{tikz}

\usepackage[group-separator={,}, group-minimum-digits=4]{siunitx}

%% file: setting_variable.tex
\newcommand{\rqa}{$RQ_1$}
\newcommand{\rqb}{$RQ_2$}
\newcommand{\rqaa}{How often are agent-generated methods deleted, and what characterizes them?}
\newcommand{\rqbb}{How accurately can method deletions be predicted?}

\newcommand{\rqA}{\rqa: \rqaa}
\newcommand{\rqB}{\rqb: \rqbb}

\newcommand{\TotalFuzzingProjectNum}{1311}
\newcommand{\TargetProjectNum}{628}

\pgfmathsetmacro{\TargetProjectRateNum}{round((\TargetProjectNum/\TotalFuzzingProjectNum)*100*10)/10}

\newcommand{\ExcludedProjectNum}{4}
\pgfmathsetmacro{\FinallyTargetProjectInt}{round(\TargetProjectNum - \ExcludedProjectNum)}



%% file: setting_fixed.tex
\definecolor{darkgreen}{rgb}{0, 0.5, 0} 
\definecolor{whitesmoke}{rgb}{0.99, 0.99, 0.99} 

\def\Underline{\setbox0\hbox\bgroup\let\\\endUnderline}
\def\endUnderline{\vphantom{y}\egroup\smash{\underline{\box0}}\\}
\def\|{\verb|}

\newcommand{\ie}{\textit{i.e.,}\xspace}
\newcommand{\eg}{\textit{e.g.,}\xspace}

\newcommand{\etal}{\xspace\textit{et al.}\xspace}

\newcommand{\approach}{\noindent\textbf{Approach. }}
\newcommand{\results}{\smallskip\noindent\textbf{Results. }}

\newcounter{findings_no}

\usepackage{listings}
\definecolor{backcolour}{rgb}{0.95,0.95,0.92}
\lstdefinelanguage{diff}{
  morecomment=**[f][\color{red}]{-},         
  morecomment=**[f][\color{darkgreen}]{+},       
  moredelim=**[is][\bfseries]{@@}{@@},
}
\definecolor{backcolour}{rgb}{0.95,0.95,0.92}
\lstdefinelanguage{commit}{ 
  breakindent = 0pt,
  numbers=none,
  backgroundcolor=\color{white},
  frame=single,
  xleftmargin=3.5em,
  numbersep=0em,
  xrightmargin=1.5em,
}

\usepackage[many]{tcolorbox}  
\tcbuselibrary{listings,breakable}
\definecolor{main}{HTML}{D0D3D4}    
\definecolor{sub}{HTML}{D0D3D4}     
\newtcolorbox{dbox}{
    enhanced,
    boxsep=3pt,       
    left=1pt,         
    right=1pt,        
    top=1pt,          
    bottom=1pt,       
    boxrule = 1pt,
    enlarge top by=1pt,
    enlarge bottom by=1pt,
  }
\tcbset{
    sharp corners,
    before skip = 0.1cm,    
    after skip = 0.01cm      
}


%% file: 1_Introduction.tex
\section{Introduction}\label{sec:introduction}

AI-assisted coding has become standard practice in modern software development. The integration of Large Language Models (LLMs) into development workflows has moved from experimental adoption to everyday use. Agentic Coding, which leverages autonomous coding agents such as GitHub Copilot~\cite{github_copilot} and Cursor~\cite{cursor}, exemplifies this shift. Developers now delegate complex tasks, including code implementation, testing, and pull request creation, to AI agents guided solely by natural language instructions.

While this shift promises substantial gains in development efficiency~\cite{DBLP:conf/chi/MozannarBFH24, DBLP:journals/corr/abs-2302-06590}, it introduces a fundamental trade-off. Agentic Coding reduces the burden on implementers but increases the burden on reviewers. Concerns regarding the quality, maintainability, and redundancy of AI-generated code are critical issues. For instance, Watanabe\etal\cite{DBLP:journals/corr/abs-2509-14745} revealed that code generated by AI agents tends to be larger than human-written code. Furthermore, Imai~\cite{DBLP:conf/icse/Imai22} showed that AI-assisted development results in more deleted code lines compared to human pair programming. These findings suggest that while Agentic Coding excels at generation, it also tends to produce excessive and unnecessary code.

Although Agentic Coding significantly reduces implementation time, the verification and review of generated code still rely on humans. The increased volume of code demands greater reviewing effort, creating a new bottleneck in the development process. Previous studies~\cite{DBLP:journals/corr/abs-2510-03029} suggest that pull requests generated by AI are likely to contain redundant methods that will eventually be deemed unnecessary, and scrutinizing all of these imposes a heavy cognitive load on human reviewers.

To the best of our knowledge, no studies have investigated the characteristics of AI-generated methods that are subsequently deleted. Predicting ``methods that will be deleted in the revisions'' would allow reviewers to prioritize their attention on essential code, thereby significantly reducing their cognitive burden.
In this study, we track the change history of code generated by Agentic Coding to quantitatively characterize (\eg lines of code, complexity, dependencies) methods deleted during review, from pull request creation to merging. We also construct a model to predict methods with a high probability of future deletion at the time of pull request creation and evaluate its effectiveness.

\smallskip\noindent\textbf{Replication Packages:} To facilitate further
studies, we provide the data and programs used in our replication package~\cite{KanPredictDeletion}.

%% file: 3_StudyDesign.tex
\begin{table*}[tb]
\centering
\caption{Detailed Definitions of Extracted Method Features}
\label{tab:features_detailed}
\small
\begin{tabularx}{\linewidth}{p{1.7cm}p{2.8cm}X}
\toprule
\textbf{Category} & \textbf{Feature Name} & \textbf{Description} \\
\midrule
\textbf{Size} 
 & \texttt{code\_loc} & Number of lines excluding comments, docstrings, and blank lines. \\
 & \texttt{char\_length} & Number of characters in the method. \\
 & \texttt{tokens} & Number of tokens extracted by Python's \texttt{tokenize} module excluding comments, indentations, and newlines. \\
 & \texttt{docstring\_words} & Number of words in the method's docstring (0 if no docstring exists). \\
 & \texttt{method\_name\_words} & Number of words in the method name, split by snake\_case and CamelCase naming conventions. \\
 \midrule
\textbf{Method Type} 
 & \texttt{is\_getter} & Boolean indicating if the name starts with \texttt{get} (case-insensitive regex match). \\
 & \texttt{is\_setter} & Boolean indicating if the name starts with \texttt{set} (case-insensitive regex match). \\
 & \texttt{is\_IsHas} & Boolean indicating if the name starts with \texttt{is} or \texttt{has} (case-insensitive regex match). \\
 & \texttt{is\_test} & Boolean indicating if "test" is included in the method name (case-insensitive). \\
 & \texttt{is\_in\_test\_code} & Boolean indicating if ``test'' is included in the file path or method name. \\
 & \texttt{is\_private} & Boolean indicating if the name starts with \texttt{\_} (excluding dunder methods). \\
 & \texttt{is\_dunder\_method} & Boolean indicating if the method is a dunder method (name surrounded by    ``\texttt{\_\_}'', e.g., \texttt{\_\_init\_\_}). \\
 & \texttt{has\_return} & Boolean indicating if the method contains a \texttt{return} statement with a value. \\
\midrule
\textbf{Contents} 
 & \texttt{param\_count} & Number of parameters (positional, keyword, \texttt{*args}, \texttt{**kwargs}). \\
 & \texttt{number\_of\_variable} & Number of variables in the method, calculated with \texttt{ast.Assign} and \texttt{ast.AnnAssign}. \\
 & \texttt{call\_expression\_count} & Number of function call nodes (\texttt{ast.Call}) within the method body. \\
 & \texttt{number\_of\_print} & Number of function calls explicitly invoking \texttt{print}. \\
 & \texttt{comment\_ratio} & Ratio of comment lines (lines starting with \texttt{\#}) to the total physical lines of code. \\
 & \texttt{cyclomatic\_complexity} & Cyclomatic complexity calculated via the \texttt{radon} library (default: 1 if calculation fails). \\
 & \texttt{halstead\_volume} & Halstead Volume calculated via the \texttt{radon} library (default: 0.0 if calculation fails). \\
 & \texttt{max\_nesting\_depth} & Maximum nesting depth of control structures (\texttt{If}, \texttt{For}, \texttt{While}, \texttt{Try}, \texttt{With}) in the AST. \\
 & \texttt{uses\_try\_except} & Boolean indicating the presence of exception handling blocks (\texttt{try-except}). \\
 & \texttt{uses\_constants} & Boolean indicating if the method uses constant-style variables. \\
\bottomrule
\end{tabularx}
\end{table*}

\section{Data Collection}\label{sec:study_design}

\subsection{Dataset}\label{sec:data_collection}
We collect methods generated by agents at the time of PR creation and analyze whether they are deleted by the time the PRs are merged. To collect PRs created by agents (Agentic-PRs), we use the AIDev dataset \cite{DBLP:journals/corr/abs-2507-15003}, which contains 33,596 PRs created by five major AI agents.

We first extracted PRs generated by agents by querying the \texttt{pull\_request} table in the dataset. We then filtered for PRs containing changes in Python code because the refactoring detection tool we use later supports only Python. 
We further filtered for merged pull requests only, as this allows us to determine whether each method was ultimately needed for production. This filtering resulted in 1,664 PRs from 197 projects. From these PRs, we extracted methods and filtered them to ensure each method was unique, identified by its file path, class name, and method name. Next, we identified three revisions from each collected pull request: the base revision (when a branch was created for the PR), the PR creation revision (when the PR was opened), and the merge revision (when the PR was integrated into the target branch, \eg main).

\subsection{Method Identification}\label{sec:method_identification}
First, we identified methods in the changed Python files of each PR. We constructed Abstract Syntax Trees (ASTs) using Python's \texttt{ast} module \cite{python_ast} to parse all methods at each revision. Specifically, we traversed \texttt{FunctionDef} and \texttt{AsyncFunctionDef} nodes, extracting class methods, top-level functions, and nested functions. Each method was uniquely identified by the combination of file path, class name, and method name.

We then identified the methods added between the base revision and the PR creation revision by comparing the ASTs of the two revisions. Similarly, we identified the methods deleted between the PR creation revision and the merge revision. Methods present in the PR creation revision but absent in the base revision were considered added, while methods present in the PR creation revision but absent in the merge revision were considered deleted.

However, between these revisions, methods, classes, and directories may have been renamed or moved. To address this, we detected refactoring operations using \texttt{ActRef}~\cite{actref_replication_2025}, a state-of-the-art refactoring detection tool for Python. It detects 15 types of refactoring operations, such as ``Rename Method'', ``Move Method'', ``Inline Method'', and ``Extract Method''. Previous studies reported that the tool achieves 78\% precision and 91\% recall~\cite{DBLP:journals/corr/abs-2505-06553}.

We applied this tool to all intermediate revisions in chronological order, both between the base revision and the PR creation revision and between the PR creation revision and the merge revision. By executing \texttt{ActRef} for each adjacent revision pair ($C_i$ and $C_{i+1}$) and tracking changes in method names, class names, and file paths step by step, we accurately determined the survival status of methods that underwent multiple refactorings or complex changes. When we detected rename or move refactoring, we did not classify these methods as added or deleted.

Based on this analysis, we label the added methods that were deleted between the PR creation revision and the merge revision as ``Deleted Methods'', while methods that persisted until the merge revision (including those that were renamed or moved) are labeled as ``Survived Methods''.


\subsection{Metrics Measurement}
To characterize the properties of methods generated by agents, we extracted 23 features from each method at the PR creation revision. Table \ref{tab:features_detailed} presents the detailed definitions of these features.
These features are organized into three categories: Size, Method Type, and Contents, based on metrics commonly used in code quality analysis~\cite{DBLP:journals/tse/BuseW10}. 
Specifically, we used the \texttt{radon} library~\cite{radon} to calculate cyclomatic complexity and Halstead volume. We extracted other features by analyzing the AST, source code tokens, and method signatures using Python's standard \texttt{ast}, \texttt{tokenize}, and \texttt{re} modules.





%% file: 4_RQ1.tex

\section{Research Questions}\label{sec:results}
\subsection*{\rqA}\label{sec:rqa}

\approach
We calculate the proportion of methods introduced at PR creation that are deleted by the time of merging. To identify characteristics that distinguish deleted methods from survived ones, we compare the distributions of the 23 extracted features between the two groups. We apply the Mann-Whitney U test to determine statistical significance and calculate Cliff's delta effect size to quantify the magnitude of these differences.

\results
We identified 12,343 methods added at PR creation. Of these, 97.4\% survived and 2.6\% were deleted (\ie 323 methods). Among the deleted methods, 1.0\% were deleted due to file deletions and 1.6\% were deleted independently. While this percentage appears small, most PRs (75.6\%) contain no revisions. Focusing on PRs with revisions (397 PRs containing 3,257 methods), 9.9\% of methods are deleted (6.0\% at the method level and 3.9\% due to file deletions). Note that the following analysis, including RQ2, excludes file-level deletions because these may not reflect the characteristics of the deleted methods themselves.

\begin{figure}[tb]
    \centering
    \includegraphics[width=0.99\linewidth]{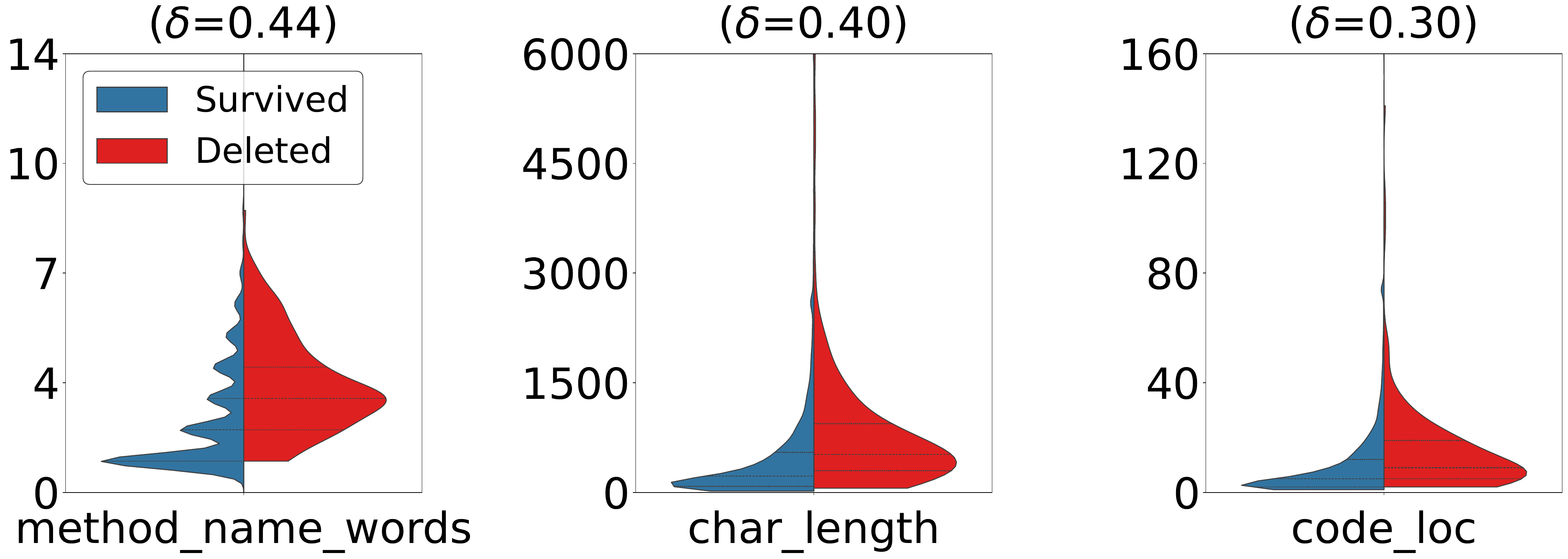}
    \caption{Feature difference (Top-3)}
    \label{fig:rq2violine}
\end{figure}

We then compared the metrics of surviving and deleted methods. Out of 23 metrics, we found statistically significant differences in 18 metrics. \autoref{fig:rq2violine} shows the top 3 features with the highest effect size, comparing ``Survived'' and ``Deleted'' methods. Additional figures are available in the online appendix.\footnote{\url{http://github.com/kan0803/predict-deletion-method/blob/main/RQ1\%20violine\_all.pdf}} 

The feature \texttt{method\_name\_words} had the highest effect size, indicating that deleted methods tend to have longer method names. For \texttt{char\_length}, deleted methods contained more characters. As shown in the figure,  \texttt{code\_loc} indicates that deleted methods tend to contain more row of code.



\smallskip
\begin{dbox}
\textbf{Answer to RQ1.}
9.9\% of methods are deleted in PRs that require revisions: 6.0\% are deleted at the method level and 3.9\% are deleted at the file level.
\end{dbox}


%% file: 4_RQ3.tex

\subsection*{\rqB}\label{sec:rqb}

\approach
We develop a binary classification model to identify methods added in PRs that will be deleted before the PRs are merged, using the features shown in Table \ref{tab:features_detailed}. We use the Random Forest (RF) algorithm because it handles heterogeneous features well, provides interpretable feature importance, and has shown strong performance in software engineering classification tasks \cite{DBLP:journals/access/DewanganRMG21}. To build the model with imbalanced labels in the dataset, we undersample the majority class (``Survived'') to match the minority class. 

\smallskip\noindent\textit{Baselines.}
We establish two baselines for comparison: random prediction and a commercial LLM. For random prediction, we assign binary labels based on the ratio of labels in the dataset. For the LLM baseline, we provide the source code of the method as input to GPT-4o. A previous study~\cite{DBLP:conf/naacl/ChenLWL25} shows that LLMs can be useful for detecting dead code. Specifically, we use the following prompt: ``\textit{Predict if the following method will be deleted (0) or survived (1) in the code review. Briefly explain the reason. The output should be in JSON.}''

\smallskip\noindent\textit{Evaluation.}
To evaluate model performance, we conducted 10-fold cross-validation and computed the median of performance metrics across folds. We used Accuracy, Precision, Recall, F1-score, and AUC as evaluation metrics. Additionally, we analyzed which features contribute to prediction by calculating the median feature importance from Random Forest.

\vspace{-3mm}
\begin{table}[h]
\small
\centering
\caption{Prediction results}
\vspace{-3mm}
\label{tab:rq3results}
\begin{tabular}{lrrr}
\toprule
evaluation          & Random & GPT-4o & Proposed\\
\midrule
AUC                   & 0.500   & 0.621           & \textbf{0.871}\\
Accuracy              & 0.498   & \textbf{0.836}   & 0.735\\
F1-Score              & 0.029   & 0.004            & \textbf{0.092}\\
Precision             & 0.015   & 0.002            & \textbf{0.049}\\
Recall                & 0.475   & 0.026            & \textbf{0.825}\\
\bottomrule
\end{tabular}
\end{table}
\results
Table \ref{tab:rq3results} shows the prediction performance of our model, the LLM, and the random approach. The proposed method achieved an AUC of 87.1\%, exceeding both GPT-4o and the random baseline. GPT-4o exhibited higher Accuracy (83.6\%) than the proposed method and the random baseline. However, GPT-4o's Recall was extremely low (2.6\%), indicating a tendency to classify most instances as the majority class, ``Survived''. In contrast, although the proposed method had low Precision (4.9\%), its Recall was high at 82.5\%, demonstrating its ability to detect a wide range of methods that are actually deleted.

Figure~\ref{fig:mismatch_example} shows an example of a method that was actually deleted. While our proposed approach correctly predicted its deletion, GPT-4o misclassified it as ``Survived''. GPT-4o justified its prediction by citing the code's readability and robustness. Specifically, the model praised the code's adherence to programming best practices, including type hints, side-effect avoidance using \texttt{.copy()}, and clear naming conventions, concluding that it was well-maintained code likely to be preserved. However, this code was deleted in reality. This instance suggests that LLMs conflate local code quality with system-level necessity.

\begin{figure}[h]
    \centering
    \begin{lstlisting}[language=Python, basicstyle=\scriptsize\ttfamily, breaklines=true, frame=single]
def transform_container_instances(instances: List[Dict[str, Any]], region: str) -> List[Dict[str, Any]]:
    transformed: List[Dict[str, Any]] = []
    for inst in instances:
        i = inst.copy()
        i["registeredAt"] = dict_date_to_epoch(i, "registeredAt")
        i["Region"] = region
        transformed.append(i)
    return transformed
    \end{lstlisting}
    \caption{Misclassified Example by GPT.}
    \label{fig:mismatch_example}
\end{figure}


Table \ref{tab:rq3importance} summarizes the top 10 feature importance values obtained from the Random Forest model.
The top-ranked features, \texttt{method\_name\_\allowbreak words}, \texttt{char\_\allowbreak length}, and \texttt{tokens}, are size-related. However, \texttt{method\_name\_\allowbreak words} exhibits the highest standard deviation (0.009), indicating variability in its predictive contribution across iterations.
This suggests that long variable names do not deterministically lead to deletion.

The contents metrics \texttt{call\_expression\_count}, \texttt{halstead\_\allowbreak volu\allowbreak me}, and \texttt{param\_count} also show relatively higher importance.
Interestingly, method type features such as \texttt{is\_private}, \texttt{is\_\allowbreak IsHas}, and \texttt{is\_getter} showed relatively low importance and did not appear in the top 10 features. 
This suggests that these features provided little utility in predicting the target class.
This implies that deletion prediction cannot rely on simple method type classifications.


\begin{table}[tb]
\centering
\caption{Feature importance rate (Top-10)}
\vspace{-2mm}
\label{tab:rq3importance}
\begin{tabular}{lrr}
\toprule
Features & Median & Std \\
\midrule
method\_name\_words & 0.16 & 0.015 \\
char\_length & 0.14 & 0.005 \\
tokens & 0.09 & 0.006 \\
code\_loc & 0.07 & 0.002 \\
call\_expression\_count & 0.07 & 0.003 \\
docstring\_words & 0.06 & 0.003 \\
comment\_ratio & 0.06 & 0.007 \\
halstead\_volume & 0.05 & 0.005 \\
param\_count & 0.05 & 0.005 \\
number\_of\_variable & 0.05 & 0.004 \\
\bottomrule
\end{tabular}
\end{table}


\smallskip
\begin{dbox}
\textbf{Answer to RQ2.}
The proposed model achieved high predictive performance with an AUC of 87.1\%, outperforming GPT-4o.
\end{dbox}
\bigskip

%% file: 6_RelatedWork.tex
\section{Related Work}\label{sec:relatedwork}

While agentic coding is expected to accelerate development through automated code generation, empirical studies suggest it can also impose a significant burden on developers. In particular, the time required to review AI-generated code has emerged as a critical concern~\cite{DBLP:journals/corr/abs-2507-09089, DBLP:conf/chi/MozannarBFH24}.
Mozannar\etal\cite{DBLP:conf/chi/MozannarBFH24} conducted a qualitative study of programmer activities and reported that ``verifying suggestions'' accounts for 22.4\% of total session time.
Becker\etal\cite{DBLP:journals/corr/abs-2507-09089} studied the impact of AI tools and found that their use actually increased development time by 19\%. They attributed this increase to the time spent reviewing and correcting the generated code.

The quality of AI-generated code is also a subject of research~\cite{DBLP:conf/icsm/AbbassiSNK25, DBLP:conf/issre/CotroneoIL25, DBLP:journals/corr/abs-2509-14745, DBLP:conf/icse/Imai22, DBLP:journals/corr/abs-2510-03029}.
Abbassi\etal\cite{DBLP:conf/icsm/AbbassiSNK25} systematically classified inefficiencies in AI-generated code, revealing that about 90\% of the generated output contained some form of inefficiency.
Cotroneo\etal\cite{DBLP:conf/issre/CotroneoIL25} identified a trend where AI-generated code frequently includes redundant or unnecessary operations.
Imai~\cite{DBLP:conf/icse/Imai22} demonstrated that while AI increases the volume of code produced, a significant portion of those lines is subsequently deleted in later stages of the development process.

Several studies have also proposed methods to detect such redundant or suboptimal code in AI-generated output~\cite{DBLP:journals/corr/abs-2504-12608, DBLP:journals/corr/abs-2407-02402, DBLP:conf/icsm/AlamRARRS25,DBLP:journals/corr/abs-2509-20491}.
Liu\etal\cite{DBLP:journals/corr/abs-2504-12608} proposed DeRep, a tool designed to detect ``repetitions'' in LLM-generated code.
Zhang\etal\cite{DBLP:journals/corr/abs-2407-02402} evaluated the effectiveness of LLMs in detecting code clones and found that LLMs perform better at identifying clones within AI-generated code than in human-written code.
Mahmoudi\etal\cite{DBLP:journals/corr/abs-2509-20491} introduced SpecDetect4AI for detecting AI-specific code smells, achieving 89\% precision and recall while being 13–129 times faster than conventional methods.

Despite these advances in identifying reviewer burdens and code quality issues, how often developers delete methods and what characteristics lead to deletion in practice remain unclear.

%% file: 7_Threat2Validity.tex
\section{Threats to Validity}\label{sec:threats_to_validity}

\noindent\textbf{Threats to internal validity:}
To accurately label methods as ``Survived'' or Deleted,'' we rely on ActRef, a Python-based refactoring detection tool. While the tool shows strong performance compared with other refactoring detection tools for Python~\cite{DBLP:conf/scam/AtwiLTKKUBL21}, we cannot entirely eliminate misclassification risks. This noise may affect label quality and introduce bias into model evaluation.

\noindent\textbf{Threats to construct validity:}
Our definition of ``Deleted'' methods is limited to the review period, from pull request creation to merging. Methods deleted after merging are not captured. Additionally, deletion does not always indicate poor code quality; methods may be removed due to requirement changes, design decisions, or consolidation with existing functionality. The features we extracted, such as cyclomatic complexity and lines of code, may not fully capture all factors influencing deletion decisions.

\noindent\textbf{Threats to external validity:}
This study exclusively targeted Python projects due to constraints of the refactoring detection tool and dataset. Python's syntax and coding conventions may differ significantly from statically typed languages such as Java or C++. Certain features, such as \texttt{is\_dunder\_method}, rely on Python-specific specifications. Consequently, our findings and prediction model may not generalize to other programming languages.

%% file: references.bib
@misc{github_copilot,
  title = {GitHub Copilot documentation},
  author = {{GitHub, Inc.}},
  year = {2025},
  url = {https://docs.github.com/en/copilot},
  note = {Accessed: December 31st, 2025}
}

@misc{cursor,
  title = {Cursor Documentation},
  author = {{Anysphere}},
  year = {2025},
  url = {https://docs.cursor.com/},
  note = {Accessed: December 31st, 2025}
}

@inproceedings{DBLP:conf/chi/MozannarBFH24,
  author       = {Hussein Mozannar and
                  Gagan Bansal and
                  Adam Fourney and
                  Eric Horvitz},
  title        = {Reading Between the Lines: Modeling User Behavior and Costs in AI-Assisted
                  Programming},
  booktitle    = {Proceedings of the 2024 {CHI} Conference on Human Factors in Computing
                  Systems, {CHI} 2024},
  pages        = {142:1--142:16},
  publisher    = {{ACM}},
  year         = {2024},
}

@article{DBLP:journals/corr/abs-2302-06590,
  author       = {Sida Peng and
                  Eirini Kalliamvakou and
                  Peter Cihon and
                  Mert Demirer},
  title        = {The Impact of {AI} on Developer Productivity: Evidence from GitHub
                  Copilot},
  journal      = {CoRR},
  volume       = {abs/2302.06590},
  year         = {2023}
}

@article{DBLP:journals/corr/abs-2509-14745,
  author       = {Miku Watanabe and
                  Hao Li and
                  Yutaro Kashiwa and
                  Brittany Reid and
                  Hajimu Iida and
                  Ahmed E. Hassan},
  title        = {On the Use of Agentic Coding: An Empirical Study of Pull Requests
                  on GitHub},
  journal      = {CoRR},
  volume       = {abs/2509.14745},
  year         = {2025}
}

@article{DBLP:journals/corr/abs-2510-03029,
  author       = {Debalina Ghosh Paul and
                  Hong Zhu and
                  Ian Bayley},
  title        = {Investigating The Smells of {LLM} Generated Code},
  journal      = {CoRR},
  volume       = {abs/2510.03029},
  year         = {2025}
}

@article{DBLP:journals/corr/abs-2505-06553,
  author       = {Siqi Wang and
                  Xing Hu and
                  Xin Xia and
                  Xinyu Wang},
  title        = {ActRef: Enhancing the Understanding of Python Code Refactoring with
                  Action-Based Analysis},
  journal      = {CoRR},
  volume       = {abs/2505.06553},
  year         = {2025}
}

@manual{python_ast,
  title = {{ast} --- {Abstract} {Syntax} {Tree}},
  author = {{Python Software Foundation}},
  organization = {Python Software Foundation},
  year = {2025}
}

@inproceedings{DBLP:conf/icse/Imai22,
  author       = {Saki Imai},
  title        = {Is GitHub Copilot a Substitute for Human Pair-programming? An Empirical
                  Study},
  booktitle    = {Proceedings of the {ACM/IEEE} 44th International Conference on Software Engineering:
                  Companion Proceedings, {ICSE} Companion 2022},
  pages        = {319--321},
  publisher    = {{ACM/IEEE}},
  year         = {2022}
}

@article{DBLP:journals/corr/abs-2507-15003,
  author       = {Hao Li and
                  Haoxiang Zhang and
                  Ahmed E. Hassan},
  title        = {The Rise of {AI} Teammates in Software Engineering {(SE)} 3.0: How
                  Autonomous Coding Agents Are Reshaping Software Engineering},
  journal      = {CoRR},
  volume       = {abs/2507.15003},
  year         = {2025},
}

@article{DBLP:journals/corr/abs-2504-12608,
  author       = {Mingwei Liu and
                  Juntao Li and
                  Ying Wang and
                  Xueying Du and
                  Zuoyu Ou and
                  Qiuyuan Chen and
                  Bingxu An and
                  Zhao Wei and
                  Yong Xu and
                  Fangming Zou and
                  Xin Peng and
                  Yiling Lou},
  title        = {Code Copycat Conundrum: Demystifying Repetition in LLM-based Code
                  Generation},
  journal      = {CoRR},
  volume       = {abs/2504.12608},
  year         = {2025}
}

@inproceedings{DBLP:conf/icsm/AlamRARRS25,
  author       = {Ajmain Inqiad Alam and
                  Palash Ranjan Roy and
                  Farouq Al{-}Omari and
                  Chanchal K. Roy and
                  Banani Roy and
                  Kevin A. Schneider},
  title        = {Are Classical Clone Detectors Good Enough for the {AI} Era?},
  booktitle    = {Proceedings of the 41st {IEEE} International Conference on Software Maintenance and Evolution,
                  {ICSME} 2025},
  pages        = {295--307},
  year         = {2025}
}

@inproceedings{DBLP:conf/naacl/ChenLWL25,
  author       = {Minyu Chen and
                  Guoqiang Li and
                  Ling{-}I Wu and
                  Ruibang Liu},
  title        = {{DCE-LLM:} Dead Code Elimination with Large Language Models},
  booktitle    = {Proceedings of the 2025 Conference of the Nations of the Americas
                  Chapter of the Association for Computational Linguistics: Human Language
                  Technologies, {NAACL} 2025 - Volume 1: Long Papers},
  pages        = {9942--9955},
  year         = {2025}
}

@misc{KanPredictDeletion,
  author       = {Kan Watanabe},
  title        = {{Replication package for: What to Cut? Predicting Unnecessary Methods in Agentic Code Generation}},
  year         = {2025},
  howpublished = {GitHub repository, \url{https://github.com/kan0803/predict-deletion-method}},
  note         = {Accessed: 2026-01-26}
}

@software{radon,
  author = {Lacchia Michele},
  title = {Radon: Code Metrics in Python},
  url = {https://github.com/rubik/radon},
  version = {6.0.1},
  date = {2023-04-12},
  year = {2025}
}

@article{DBLP:journals/tse/BuseW10,
  author       = {Raymond P. L. Buse and
                  Westley Weimer},
  title        = {Learning a Metric for Code Readability},
  journal      = {{IEEE} Transactions on Software Engineering, {TSE}},
  volume       = {36},
  number       = {4},
  pages        = {546--558},
  year         = {2010}
}

@article{DBLP:journals/corr/abs-2507-09089,
  author       = {Joel Becker and
                  Nate Rush and
                  Elizabeth Barnes and
                  David Rein},
  title        = {Measuring the Impact of Early-2025 {AI} on Experienced Open-Source
                  Developer Productivity},
  journal      = {CoRR},
  volume       = {abs/2507.09089},
  year         = {2025}
}

@inproceedings{DBLP:conf/icsm/AbbassiSNK25,
  author       = {Altaf Allah Abbassi and
                  L{\'{e}}uson M. P. da Silva and
                  Amin Nikanjam and
                  Foutse Khomh},
  title        = {A Taxonomy of Inefficiencies in LLM-Generated Python Code},
  booktitle    = {Proceedings of the 41st {IEEE} International Conference on Software Maintenance and Evolution,
                  {ICSME} 2025},
  pages        = {393--404},
  publisher    = {{IEEE}},
  year         = {2025}
}

@inproceedings{DBLP:conf/issre/CotroneoIL25,
  author       = {Domenico Cotroneo and
                  Cristina Improta and
                  Pietro Liguori},
  title        = {Human-Written vs. AI-Generated Code: {A} Large-Scale Study of Defects,
                  Vulnerabilities, and Complexity},
  booktitle    = {Proceedings of the 36th {IEEE} International Symposium on Software Reliability Engineering,
                  {ISSRE} 2025},
  pages        = {252--263},
  publisher    = {{IEEE}},
  year         = {2025}
}

@article{DBLP:journals/corr/abs-2407-02402,
  author       = {Zixian Zhang and
                  Takfarinas Saber},
  title        = {Assessing the Code Clone Detection Capability of Large Language Models},
  journal      = {CoRR},
  volume       = {abs/2407.02402},
  year         = {2024}
}

@article{DBLP:journals/corr/abs-2509-20491,
  author       = {Brahim Mahmoudi and
                  Naouel Moha and
                  Quentin Stievenert and
                  Florent Avellaneda},
  title        = {AI-Specific Code Smells: From Specification to Detection},
  journal      = {CoRR},
  volume       = {abs/2509.20491},
  year         = {2025},
}

@inproceedings{DBLP:conf/scam/AtwiLTKKUBL21,
  author       = {Hassan Atwi and
                  Bin Lin and
                  Nikolaos Tsantalis and
                  Yutaro Kashiwa and
                  Yasutaka Kamei and
                  Naoyasu Ubayashi and
                  Gabriele Bavota and
                  Michele Lanza},
  title        = {{PYREF:} Refactoring Detection in Python Projects},
  booktitle    = {Proceedings of the 21st {IEEE} International Working Conference on Source Code Analysis
                  and Manipulation, {SCAM} 2021},
  pages        = {136--141},
  year         = {2021}
}

@article{DBLP:journals/access/DewanganRMG21,
  author       = {Seema Dewangan and
                  Rajwant Singh Rao and
                  Alok Mishra and
                  Manjari Gupta},
  title        = {A Novel Approach for Code Smell Detection: An Empirical Study},
  journal      = {{IEEE} Access},
  volume       = {9},
  pages        = {162869--162883},
  year         = {2021}
}

@misc{actref_replication_2025,
  author       = {Siqi Wang and Xing Hu and Xin Xia and Xinyu Wang},
  url          = {https://figshare.com/s/984c7a39266137e29c37},
  note         = {Accessed: December 31st, 2025},
  year         = {2025}
}
